\documentstyle{article}
\begin{document}
\begin{center}
\Large\textbf{On statistical mechanics in noncommutative spaces.}
\end{center}
\begin{center}
\textbf{S. A. Alavi}

\textit{Department of Physics, Sabzevar university of Tarbiat Moallem, Sabzevar, P. O. Box 397,
 Iran\\
  and \\
 Sabzevar House of Physics, Sabzevar, Iran.}\\
\end{center}

\begin{abstract}
We study the formulation of quantum statistical mechanics in noncommutative spaces. We construct microcanonical and canonical ensemble theory in noncommutative spaces. We consider for illustration some basic and important examples in the framework of noncommutative statistical mechanics : (i). An electron in a magnetic field. (ii). A free particle in a box. (iii). A linear harmonic oscillator.
 
\end{abstract}

 
\textbf{Pacs}: 02.40.Gh. 05.30.-d\\

\textbf{Introduction.}\\

To study quantum mechanical systems composed of indistiguishale entities, as most physical systems are, one finds that it is advisable to rewrite the ensemble theory in a language that is more natural to a quantum-mechanical treatment, namely the language of the operators and the wave functions[1]. Once we set out to study these systems in detail, we encounter a stream of new and altogether different physical concepts. In particular, we find that the behavior of even a noninteracting system, such as the ideal gas, departs considerably from the pattern set by the so-called calssical treatments. In the presence of interactions the pattern becomes still more complicated. \\
Recently there have been notable studies on the formulation and possible experimental consequences of extensions of the usual physical theories in the noncommutative spaces[2]. The study on noncommutative spaces is much important for understanding phenomena at short distances beyond the present test of different physical theories. For a manifold parameterized by the coordinates $x_{i}$, the noncommutative relations can be written as :

\begin{equation}
[\hat{x}_{i},\hat{x}_{j}]=i\theta_{ij}\hspace{1.cm}[\hat{x}_{i},\hat{p}_{j}]=i\delta
_{ij}\hspace{1.cm} [\hat{p}_{i},\hat{p}_{j}]=0 ,
\end{equation}
where $\theta_{ij}$ is an antisymmetric tensor which can be defined as $\theta_{ij}=\frac{1}{2}\epsilon_{ijk}\theta_{k}$. \\
In this paper we study the formulation of quantum statistics, namely the quantum-mechanical ensemble theory, the density matrix, etc., in a noncommutative space and the new features that arise. 

\textbf{Perturbation aspects of noncommutative dynamics.}\\

NCQM is formulated in the same way as the standard quantum mechanics SQM
(quantum mechanics in commutative spaces), that is in terms of the
same dynamical variables represented by operators in a Hilbert
space and a state vector that evolves according to the
Schroedinger equation :
\begin{equation}
i\frac{d}{dt}|\psi>=H_{nc}|\psi> ,
\end{equation}
we have taken in to account $\hbar=1$. $H_{nc}\equiv H_{\theta}$
denotes the Hamiltonian for a given system in the noncommutative
space. In the literatures two
approaches have been considered  for constructing the NCQM :\\
a) $H_{\theta}=H$, so that the only difference between SQM and
NCQM is the presence of a nonzero $\theta$ in the commutator of
the position operators i.e. Eq.(1). \\
b) By deriving the Hamiltonian from the moyal analog of the
standard Schroedinger equation :
\begin{equation}
i\frac{\partial}{\partial t}\psi(x,t)=H(p=\frac{1}{i}\nabla,x)\ast
\psi (x,t)\equiv H_{\theta}\psi(x,t) ,
\end{equation}
where $H(p,x)$ is the same Hamiltonian as in the standard theory,
and as we observe the $\theta$ - dependence enters now through the
star product [3]. In [4], it has been shown that these two
approaches lead to the same physical theory. Since the noncommutativity parameter if it is non-zero, should be very small compared to the length scales of the system, one can always treat the noncommutativity effects as some perturbations of the commutative counterpart.\\ 
For the Hamiltonian
of the type :
\begin{equation}
H(\hat{p},\hat{x})=\frac{\hat{p}^{2}}{2m}+V(\hat{x}) .
\end{equation}
The modified Hamiltonian $H_{\theta}$ can be obtained by a shift
in the argument of the potential [5] :
\begin{equation}
x_{i}=\hat{x}_{i}+\frac{1}{2}\theta_{ij}\hat{p}_{j}\hspace{2.cm}\hat{p}_{i}=p_{i} .
\end{equation}
which lead to
\begin{equation}
H_{\theta}=\frac{p^{2}}{2m}+V(x_{i}-\frac{1}{2}\theta_{ij}p_{j}) .
\end{equation}
The variables $x_{i}$ and $p_{i}$ now, satisfy in the same
commutation relations as the usual case :
\begin{equation}
[x_{i},x_{j}]=[p_{i},p_{j}]=0\hspace{1.cm}[x_{i},p_{j}]=\delta_{ij} .
\end{equation}

Now we discuss the perturbation aspects of
noncommutative dynamics. Using
\begin{equation}
U(x+\Delta x)=U(x)+\sum_{n=1}^{\infty}  \frac{U^{(n)}(x)}{n!} (\Delta x)^{n} ,
\end{equation}
and Eq.(6) we have :
\begin{equation}
H_{nc}=\frac{p^{2}}{2m}+V(x_{i})+\sum_{n=1}^{\infty}  
\frac{V^{(n)}(x_{i})}{n!} (\Delta x_{i})^{n} ,
\end{equation}
where $\Delta x_{i}=-\frac{1}{2}\theta \epsilon_{ij}p_{j} $ and
$H=\frac{p^{2}}{2m}+V(x)$ is the Hamiltonian in ordinary(commutative) space.
To the first order we have :
\begin{equation}
H_{nc}\equiv H_{\theta}=\frac{p^{2}}{2m}+V(x_{i})+\Delta x_{i} \frac{\partial
V}{\partial x_{i}}=H+\Delta x_{i}\frac{\partial
V}{\partial x_{i}}=H+\theta H_{I} .
\end{equation}

We can use perturbation theory to
obtain the eigenvalues and eigenfunctions of $H_{nc}$ :
\begin{equation}
E_{n}=E^{0}_{n}+\Delta E^{0}_{n}=E^{0}_{n}+\theta E^{(1)}_{n}+\theta^{2} 
E^{(2)}_{n}+...\hspace{.3cm} .
\end{equation}
\begin{equation}
\hat{\phi}_{n}=\phi_{n}+ \sum_{k\neq n}C_{nk}(\theta)\phi_{k} .
\end{equation}
where :
\begin{equation}
C_{nk}(\theta)=\theta C^{(1)}_{nk}+\theta^{2} C^{(2)}_{nk}+...\hspace{.3cm} .
\end{equation}
To the first order in perturbation theory we have :
\begin{equation}
\theta E^{(1)}_{n}= <\phi_{n}|\theta H_{I} |\phi_{n}> ,
\end{equation}
\begin{equation}
\hat{\phi}_{n}=\phi_{n}+\theta \sum_{k\neq n}C^{(1)}_{nk}\phi_{k} ,
\end{equation}
\begin{equation}
\theta C^{(1)}_{nk}=\frac{<\phi_{k}|\theta
H_{I}|\phi_{n}>}{E^{0}_{n}-E_{k}^{0}} ,
\end{equation}
where $E^{0}_{n}$ and $\phi_{n}$ are the $n$th eigenvalue and
eigenfunction of the Hamiltonian $H$. $E_{n}$ and $\hat{\phi}_{n}$ are
the $n$th eigenvalue and eigenfunction of $H_{nc}$.\\

\textbf{The density operator in noncommutative spaces.}\\

Using the orthonormal functions $\hat{\phi}_{n}$, an arbitrary wave function in a noncommutative space can be written as :
\begin{equation}
\hat{\psi^{k}}(t)=\sum_{n} \hat{a}_{n}^{k}(t)\hat{\phi}_{n}
\end{equation}
where :
\begin{equation}
\hat{a}_{n}^{k}(t)=\int\hat{\phi}_{n}^{*}\hat{\psi^{k}}(t)d\tau
\end{equation}

The time variation of these coefficients will given by :\\

$i\hbar\frac{\partial}{\partial t}\hat{a}_{n}^{k}=i\hbar \int\hat{\phi}_{n}^{*}\frac{\partial}{\partial t}\hat{\psi^{k}}(t)d\tau=$
\begin{equation}
\int\hat{\phi}_{n}^{*}\hat{H}\hat{\psi^{k}}(t)d\tau=\int\hat{\phi}_{n}^{*}\hat{H}{\sum_{m} \hat{a}_{n}^{k}(t)\hat{\phi}_{m}}d\tau=\sum_{m} \hat{H}_{nm}a^{k}_{m}(t)
\end{equation}
where $\hat{H}_{nm}=\int\hat{\phi}_{n}^{*}\hat{H}\hat{\phi}_{m}$. 
We now introduce the density operator $\hat{\rho}(t)$, in a noncommutative space by the matrix elements :
\begin{equation}
\hat{\rho}_{nm}(t)=\frac{1}{n}\sum^{n}_{k=1}[\hat{a}^{k}_{m}(t)\hat{a}^{k*}_{n}(t)]
\end{equation}

Clearly the matrix element $\hat{\rho}_{nm}(t)$, is the ensemble average of the quantity $a_{m}(t)a^{*}_{n}(t)$, which as a rule varies from member to member in the ensemble.\\
We shall now determine the equation of motion for the density matrix $\hat{\rho}_{mn}(t)$ :\\
\begin{equation}
i\hbar\frac{\partial}{\partial t}\hat{\rho}_{mn}(t)=\frac{1}{n}\sum^{n}_{k=1}{i\hbar[\frac{\partial}{\partial t}\hat{a}^{k}_{m}(t)]\hat{a}^{k*}_{n}(t)+i\hbar[\frac{\partial}{\partial t}\hat{a}^{k*}_{n}(t)]\hat{a}^{k}_{m}(t)}
\end{equation}
It can be written in the following form :
\begin{equation}
\frac{1}{n}\sum^{n}_{k=1}[\sum_{\ell}{\hat{H}}_{m\ell}\hat{a}^{k}_{\ell}(t)]\hat{a}^{k*}_{n}(t)-\sum^{n}_{k=1}\sum_{\ell}[{\hat{H}}^{*}_{n\ell}\hat{a}^{k*}_{\ell}(t)]\hat{a}^{k}_{m}(t)=
(\hat{H}\hat{\rho}-\hat{\rho}\hat{H})_{mn}
\end{equation}
Using the commutator notation, it can be written as :
\begin{equation} 
i\hbar \dot{\hat{\rho}}=[\hat{H},\hat{\rho}]
\end{equation}
Now, we consider the expectation value of a physicsl quantity $G$, in a noncommutative space which is dynamically represented by an operator $\hat{G}$. This will naturally be determined by the double averaging process :
\begin{equation} 
<\hat{G}>=\frac{1}{n}\sum^{n}_{k=1}\int \hat{\psi}^{k*}\hat{G}\hat{\psi}^{k}d\tau
\end{equation}
or :
\begin{equation} 
<\hat{G}>=\frac{1}{n}\sum^{n}_{k=1}\sum_{m,n}\hat{a}^{k*}_{n}\hat{a}^{k}_{m}\hat{G}_{nm}
\end{equation}
where :
\begin{equation} 
\hat{G}_{nm}=\int\hat{\phi}_{n}^{*}\hat{G}\hat{\phi}_{m}d\tau
\end{equation}
Introducing the density matrix $\hat{\rho}$, it takes a particularly neat form :
\begin{equation} 
<\hat{G}>=\sum_{m,n}\hat{\rho}_{n}\hat{G}_{nm}=\sum_{m}(\hat{\rho}\hat{G})_{mm}=Tr(\hat{\rho}\hat{G})      
\end{equation}
We note that if the original wave functions $\hat{\psi}^{k}$, were not normalized then the expectation value $<\hat{G}>$ would be given by the formula 
\begin{equation} 
<\hat{G}>=\frac{Tr(\hat{\rho}\hat{G})}{Tr(\hat{\rho)}}
\end{equation}
The interesting point is that the equations (23) and (28) are the same as the commutative case with quantities replaced by their noncommutative counterparts.\\

\textbf{Statistics of the various ensembles.}\\

\textit{The microcanonical  ensemble.}\\

The construction of the microcanonical ensemble is based on the premise that the systems constituting the ensemble 
are characterized by a fixed number of particles N, a fixed volume V and an energy lying within the interval 
$(E-\frac{\Delta}{2},E+\frac{\Delta}{2})$, where $\Delta<<E$. The total number of distinct microstates accessible to a system is then denoted by the symbole $\Gamma(N,V,E;\Delta)$ and by assumption, any of these microstates is just as likely to occur as any other.\\
Accordingly, the density matrix $\hat{\rho}_{mn}$(which in the energy representation must be a diagonal matrix) will be of the form $\frac{1}{\Gamma}$, for each of the accessible states and $0$, for all other states.\\
We note that $\hat{\rho}$ is independent of energy (energy eigenstates) and the volume V, so it does not depend on space coordinates. It means that the noncommutativity of space has no effects on $\hat{\rho}$ in microcanonical ensemble. \\
The dynamics of the system determined by the expression for its entropy, which in turn is given by :
\begin{equation} 
S=k Ln\Gamma 
\end{equation} 
which as mentioned above remains unchanged in noncommutative spaces.\\

\textit{The canonical  ensemble.}\\

In this ensemble the macrostate of a member system is defined through the parameters N, V and T; the energy E now becomes a variable quantity. The probability that a system, chosen at random from the ensemble, possesses an energy E, is determined by the Boltzmann factor $exp(-\beta E)$, where $\beta=\frac{1}{kT}$. The density matrix in the energy representation is therefore takes as :
\begin{equation} 
\hat{\rho}_{mn}=\hat{\rho}_{n}\delta_{mn} 
\end{equation}  
where :
\begin{equation} 
\hat{\rho}_{n}=c e^{-\beta E_{n}};\hspace{1.cm} n=0,1,2,...
\end{equation} 

Here $E_{n}$ are the energy eigenvalues in noncommutative space(Eq.11), and the constant $c$ is given by :
\begin{equation} 
c=\frac{1}{\sum_{n}exp(-\beta E_{n})}=\frac{1}{\hat{Q}_{N}(\beta)} 
\end{equation}
where $\hat{Q}_{N}(\beta)$, is the partition function of the system in noncommutative space. The density operator in the canonical ensemble may be written as :

$\hat{\rho}=\sum_{n}\left|\hat{\phi_{n}}\right\rangle \frac{1}{\hat{Q}_{N}(\beta)}e^{-\beta E_{n}}\left\langle \hat{\phi_{n}}\right|=$
\begin{equation} 
\frac{1}{\hat{Q}_{N}(\beta)}e^{-\beta \hat{H}}\sum_{n}\left|\hat{\phi_{n}}\right\rangle\left\langle \hat{\phi_{n}}\right|=\frac{e^{-\beta \hat{H}}}{Tr(e^{-\beta \hat{H}})}
\end{equation} 
Then the expectation value of a physical quantity $G$, in a noncommutative space is given by :
\begin{equation} 
<\hat{G}>_{N}=Tr(\hat{\rho}\hat{G})=\frac{Tr(\hat{G}e^{-\beta \hat{H}})}{Tr(e^{-\beta \hat{H}})} 
\end{equation}  

\textbf{Examples.}\\

\textit{(i). An electron in a magnetic field.}\\

The Hamiltonian of the system has the following form:
\begin{equation} 
\hat{H}=-\mu_{B}(\vec{\sigma}\cdot\vec{B} )
\end{equation}  
where $\mu_{B}=\frac{e\hbar}{2mc}$. The Hamiltonian is space independent, so there is no corrections due to the noncommutativity of space on the statistical (Thermodynamical) properties of this system.\\

\textit{(ii). A free particle in a box.}\\

 Let us consider the motion of a particle with charge e and mass m in the presence of a magnetic field produced by a vector potential $\vec{A}$. The lagrangian is as follows :
\begin{equation}
L=\frac{1}{2}m V^{2}+\frac{e}{c}\vec{A}\cdot\vec{V}-V(x,y)
\end{equation}
where $V_{i}=h_{i}\dot{q}_{i}$(no summation, $i=1,2,3$), are the components of the  velocity of the particle and $h_{i} (i=1,2,3)$ are the scale factors. $V(x,y)$ describes aditional interactions(impurities). For the case of a free particle $V(x,y)=0$. In the absence of  $V$  the quantum spectrum consists of the well-known Landau levels. In the strong magnetic field limit only the lowest Landau level is relevant. But the large B limit corresponds to small m, so setting the mass to zero effectively projects onto the lowest Landau level. In the chosen gauge $\vec{A}=(0,h_{1}q_{1}B)$ and in that limit, the Lagrangian(36) takes the following form :
\begin{equation}
L^{\prime}=\frac{e}{c}Bh_{1}h_{2}q_{1}\dot{q}_{2}-V(x,y)
\end{equation}
which is of the form $p\dot{q}-H(p,q)$,and suggests that $\frac{e}{c}Bh_{1}q_{1}$ and $ h_{2}q_{2}$ are canonical conjugates, so we have:
\begin{equation}
[h_{1}q_{1},h_{2}q_{2}]=-i\frac{\hbar c}{eB}
\end{equation}
which can be written in general form :
\begin{equation}
[h_{i}q_{i},h_{j}q_{j}]=i\theta_{ij}
\end{equation}
which is the fundamental space-space noncommutativity relation in a general noncommuting curvilinear coordinates. The cartesian, circular cylindrical and spherical polar coordinates are three special cases [6].\\
So a free particle in a noncommutative space is equal to a particle in commutative space but in the presence of a magnetic field. The Hamiltonian of a particle in a magnetic field is :

\begin{equation}
H=\frac{1}{2m}(\vec{P}-\frac{q}{c}\vec{A})^{2}
\end{equation}
On the other hand let us introduce the noncommutativity to momentums instead of space coordinates :
\begin{equation}
[\hat{x}_{i},\hat{x}_{j}]=0\hspace{1.cm}[\hat{x}_{i},\hat{p}_{j}]=i\delta
_{ij}\hspace{1.cm} [\hat{p}_{i},\hat{p}_{j}]=i\theta_{ij} ,
\end{equation}
one can easily show that there is a transformation :
\begin{equation}
\hat{p}_{i}=p_{i}+\frac{1}{2}\theta_{ij}x_{j}\hspace{1.cm}\hat{x}_{i}=x_{i}\
\end{equation}
where the new variables $p_{i}$ and $x_{i}$ satisfy the standard commutation relations (7). We note that (42) is the same as $p_{i}\rightarrow p_{i}-\frac{A_{i}}{c}$ with $A_{i}=-\frac{1}{2}\theta_{ij}x_{j}$.\\
Now the Hamiltinian of a free particle in a noncommutative space is :\\

$H=\frac{1}{2m}\hat{p}^{2}=$
\begin{equation}
\left(p_{i}+\frac{1}{2}\theta{ij}x_{j}\right)^{2}=p_{i}^{2}+\theta_{ij}p_{i}x_{j}+O(\theta^{2})=p_{i}^{2}-\frac{1}{2}\vec{L}\cdot\vec{\theta}+O(\theta^{2})
\end{equation} 
where $L_{k}=\epsilon_{ijk}x_{i}p_{j}$. Since for a free particle $\vec{L}=0$, so to the first order there is no corrections on the Hamiltonian and therefore there is no corrections  due to noncommutativity of space on the statistical(Thermodynamical) properties of  this system.\\

\textit{(iii). A harmonic oscillator.}\\

The case of a linear harmonic oscillator is irrelevance, becauase there is only one space variable. In the case of harmonic oscillator in higher dimensions for instance spherical harmonic oscillator, the corrections on the Hamiltonian due to noncommutativity of space is given by :

\begin{equation}
H_{I}=\frac{\partial V}{\partial x_{i}}\Delta x_{i}=-\frac{1}{2}\theta_{ij}\frac{\partial V}{\partial x_{i}}p_{j}
\end{equation}
where :
\begin{equation}
V=\frac{1}{2}m\omega^{2}\sum_{i=1}^{3}x_{i}^{2}
\end{equation}
We put $\theta_{3}=\theta$ and the rest of the $\theta$ components to zero, which can be done by a rotation or by a redifinition of coordinates. So we have :
\begin{equation}
H_{I}=-\frac{1}{2}\theta_{ij}(m\omega^{2}x_{i})p_{j}=\frac{1}{4}m\omega^{2}\epsilon_{ijk}x_{i}p_{j}\theta_{k}=-\frac{1}{4}\vec{L}\cdot\vec{\theta}=\frac{1}{4}L_{z}\theta
\end{equation}
For a spherical harmonic oscillator the unperturbed(commutative) eigenfunctions are given by :
\begin{equation}\phi_{nlm}(r\theta\phi)\propto \frac{r^{\ell+1}e^{\frac{-\alpha r^{2}}{2}}L_{n}^{\ell+\frac{1}{2}}(\alpha r^{2})Y_{\ell m}(\theta\phi)}{r}
\end{equation}
where $L_{n}^{\ell+\frac{1}{2}}(\alpha r^{2})$ are Lagure's functions. Using Eqs.(14) and (31), one can easily derive the enery eigenvalues in noncommutative space and so the density operator $\hat{\rho}$ and the partition function $\hat{Q}_{n}(\beta)=\sum_{n}e^{-\beta E_{n}}$. The thermodynamical properties of the system can be done straightforwardly using partition function. We have :
\begin{equation}
E^{1}_{n}=\left\langle \phi_{nlm}\left|H_{I}\right|\phi_{nlm}\right\rangle=\frac{1}{4}m\theta
\end{equation}
So :
\begin{equation}
\hat{Q}_{N}(\beta)=e^{-\frac{\beta}{4}m\theta}Q_{N}(\beta)
\end{equation}
The Helmholtz free energy is given by :
\begin{equation}
\hat{A}\frac{k+\beta}{4}m\theta+A
\end{equation}
Whence we obtain :
\begin{equation}
\hat{S}=-\frac{\partial\hat{A}}{\partial T}=\frac{k\beta}{4}m\theta+S
\end{equation}
\begin{equation}
\hat{U}=\frac{k+\beta}{2}m\theta+U
\end{equation}
\begin{equation}
\hat{C}=\frac{k\beta}{2}m\theta+C
\end{equation}
where $\hat{S}$, $\hat{U}$ and $\hat{C}$ are the entropy, the enternal energy and the specific heat of the system in noncommutative space and S, U and C are their counterpart in the commutative case.\\

\textbf{References.}\\
1. R. K. Pathria, Statistical mechanics, Butterworth-Heinemann, 1996.\\ 
2. See for example the reviewes : R. Szabo, Class. Quant. Grav. 23(2006)R199-R242 and Phys. Rept. 378(2003)207-299, and the references there in.\\
3. L. Mezincescu, hep-th/0007076.\\
4. O. Espinosa, hep-th/0206066.\\
5. M. Chaichian, M. M. Sheikh-Jabbari and A. Tureanu, Phys. Rev. Lett 86 (2001)2716\\
6. R. Jackiw, S. Y. Pi and Polychronakos, Ann. Phys.(N.Y) 301(2002) 157. R. Jackiw, Annals Henri Poincare 4S2(2003)S913.  S. A. Alavi, Chin. Phys. Lett. 23(2006)2637.
\end{document}